\documentclass[12pt]{article}
\usepackage{epsfig}
\begin{document}

\title{Comment on ``Formation of primordial black holes by cosmic strings"}
\author{R.N. Hansen\thanks{Electronic address: rnh@fysik.ou.dk}, \,
M. Christensen\thanks{Electronic address: mc@bose.fys.ou.dk} \, and
A.L. Larsen\thanks{Electronic address:  all@fysik.ou.dk}}
\date{\today}
\maketitle
\noindent
\centerline{\em Physics Department, University of Odense, }\\
\centerline{\em Campusvej 55, 5230 Odense M, Denmark}
\vskip 2cm
\begin{abstract}\baselineskip=1.5em
We show that in a pioneering paper by Polnarev and Zembowicz, some
conclusions concerning the characteristics of the Turok-strings are
generally not correct. In addition we show that the probability of
string collapse given there, is off by a large prefactor ($\sim 10^3$).
\end{abstract}

\newpage
In one of the pioneering and often cited papers on the probability of
cosmic string collapse \cite{pol}, Polnarev and Zembowicz analyzed
the 2-parameter Turok-strings
\cite{turok}:
\begin{eqnarray}
X(\tau,\sigma) & = & \frac{A}{2}~\left[{(1-\alpha)\sin (\sigma -\tau)
+ \frac{\alpha}{3}\sin
3 (\sigma - \tau)+ \sin (\sigma +\tau) }\right] \nonumber\\
Y(\tau,\sigma) & = & \frac{A}{2}~\left[{(1-\alpha)\cos (\sigma -\tau)
+ \frac{\alpha}{3}\cos
3 (\sigma - \tau)+ (1-2 \beta )
\cos (\sigma +\tau) }\right] \nonumber \\
Z(\tau,\sigma) & = & \frac{A}{2}~\left[{2\sqrt{\alpha(1-\alpha)}\cos
(\sigma -
\tau) +
2\sqrt{\beta(1-\beta)}\cos (\sigma + \tau)  }\right]
\label{turokstring}
\end{eqnarray}
(We included a dimensionfull parameter $A$ to keep $\tau$ and
$\sigma$ dimensionless). \\

It was concluded \cite{pol},
among other things, that:
\begin{itemize}
\item The strings have their minimal size $R$ at
\begin{equation}
\tau = \frac{\pi}{2}
\label{polz1}
\end{equation}
\item For generic parameters $(\alpha, \beta)$:
\begin{equation}
\frac {R^2}{A^2} = \left(\sqrt{\alpha(1-\alpha)} -
\sqrt{\beta(1-\beta)}\;\right)^2 +
\left(\frac{\alpha}{3}-\beta\right)^2
\label{polz2}
\end{equation}
\end{itemize}\vskip 6pt
We now give two simple explicit examples showing that the two
conclusions
(\ref{polz1}), (\ref{polz2}) cannot generally be correct.\\

\noindent {\bf A}. Consider first the case $\alpha = 1,\; \beta = 0$. 
Besides
$Z = 0,$ this corresponds to:
\begin{eqnarray}
X(\tau,\sigma) = \frac{A}{2} \left[{\frac{1}{3}\sin 3 (\sigma -\tau)
+ \sin(\sigma +
\tau)}\right] \nonumber\\
Y(\tau,\sigma) = \frac{A}{2} \left[{\frac{1}{3}\cos 3 (\sigma -\tau)
+ \cos(\sigma +
\tau)}\right]
\end{eqnarray}
This is in fact a rigidly rotating string:
\begin{equation}
\left(
\begin{array}{cc}
{\mbox{X}}\left( \tau, \sigma \right) \\
{\mbox{Y}}\left( \tau, \sigma \right) \\
\end{array}
\right)
=
\left(
\begin{array}{cc}
\cos(3\tau)&\sin(3\tau)\\
-\sin(3\tau)&\cos(3\tau)\\
\end{array}
\right)
\left(
\begin{array}{cc}
\mbox{X}\left(0,\tilde{\sigma}\right)\\
\mbox{Y}\left(0,\tilde{\sigma}\right)\\
\end{array}
\right)
\end{equation}
where $\tilde{\sigma} \equiv \sigma - 2\tau$. It follows that the
minimal
string size $R$ (the radius
of the minimal sphere that can ever enclose the string completely) is
independent
of time. Thus it can be computed at any time, say $\tau = 0$:
\begin{equation}
R = \begin{array}{c}
\mbox{Maximum} \\ \sigma \in [0,2\pi ]
\end{array} \left[
\sqrt{X^2(0,\sigma ) + Y^2(0,\sigma )} \, \right] 
= \frac{2A}{3} 
\end{equation}
Notice that the minimal sphere is found by maximization over
$\sigma$. Thus the result
(\ref{polz2}) is not correct in this case. In fact, it gives the
$\underline{minimal}$ distance
from origo to the string
(namely $A/3$), but to completely enclose the string, one needs a
sphere
with radius
corresponding to the
$\underline{maximal}$ distance (namely $2A/3$).\\

\noindent {\bf B}. Now consider the case $\alpha = 1/2,\;\beta = 1$. 
Let us consider the distance
from origo to the string as a function of $\sigma$ at two different
times,
namely $\tau = 0$
and $\tau =\pi/2$. It is straightforward to show that
\begin{eqnarray}
\begin{array}{c}
\mbox{Maximum} \\ \sigma \in [0,2\pi ] \end{array} \:
\left[ \sqrt{X^2(0,\sigma ) + Y^2(0,\sigma ) + Z^2(0,\sigma )} \: \right]
\: < \nonumber \\
\begin{array}{c}
\mbox{Maximum} \\ \sigma \in [0,2\pi ] \end{array} \:
\left[ \sqrt{X^2(\pi /2,\sigma ) 
+ Y^2(\pi /2,\sigma ) + Z^2(\pi /2,\sigma )} \: \right]
\end{eqnarray}
Thus the string does not have its minimal size at $\tau = \pi/2$;
at $\tau = 0$ it can be enclosed in a much smaller sphere. More
precisely, at
$\tau = 0$, the string can be enclosed in a sphere of radius
$\sqrt{155/288} \, A$
while at $\tau = \pi/2$, a sphere of radius $\sqrt{17/18} \, A$ is
needed. Therefore, the result (\ref{polz1}) is not correct in this
case.\\

On the other hand, for some other particular examples, it seemed that
the conclusions (2)-(3) were indeed correct. Thus to clarify the
situation, we did a complete re-analysis of the problem (see
\cite{rene} for the details) using both analytical and numerical
methods. This led to a precise classification of the
Turok-strings,
and a subsequent subdivision into 3 different 
families (see Fig. 1): \\

\noindent
{\bf I}. These strings have their minimal size at $\tau =\pi/2$. That
is,
starting
from their original size at $\tau$ = 0, they generally contract to
their
minimal size at
$\tau =\pi/2$, and then generally expand back to their original size
at
$\tau = \pi$.\\

\noindent
{\bf II}. These strings start from their minimal size at $\tau$ = 0.
Then they
generally expand towards their maximal size and then recontract
towards
their
minimal size at $\tau = \pi$.\\

\noindent
{\bf III}. These strings have their minimal size at two values of
$\tau$
symmetrically around
$\pi/2$. That is, they first generally contract and reach the minimal
size
at some
$\tau_0\in\left[0;
\pi/2\right]$. Then they expand for a while, and then recontract and
reach
the minimal size
again at $\tau =\pi - \tau_0$. Then they expand again towards the
original
size
at $\tau = \pi$. In this family of strings, the value of $\tau_0$
depends on
$(\alpha , \beta)$.\\
\\
Then by comparison, we see that the conclusion (2)   is correct in
the region
{\bf I} of parameter-space,
but incorrect in regions {\bf II} and {\bf III}. \\

As for the conclusion (3), let us restrict ourselves to the region
{\bf I} of parameter-space. This is the most relevant region for
string collapse since it includes the circular string
($\alpha=\beta=0$), and string collapse is only to be expected for
low angular momentum near-circular strings.
In any case, in the region {\bf I}, it is easy to derive the exact
analytical expression for the minimal string size \cite{rene}:
\begin{equation}
{R}^2 = \mbox{Max} \left( R^{2}_1,\,R^{2}_2\right)
\label{collapse}
\end{equation}
where
\begin{equation}
\frac{R^{2}_1}{A^2} = \frac{4{\alpha}^2}{9}
\label{rone}
\end{equation}
and
\begin{equation}
\frac{R^{2}_2}{A^2} = \left(\sqrt{\alpha\left(1-\alpha\right)} -
\sqrt{\beta\left(1-\beta\right)}\right)^2 + \left(\frac{\alpha}{3} -
\beta\right)^2
\label{rtwo}
\end{equation}
Notice that Eq. (\ref{rtwo}) is precisely the result (\ref{polz2})
of Polnarev and Zembowicz
\cite{pol}. However, in Ref. \cite{pol}, the other solution
(\ref{rone})
was completely
missed, and this is actually the relevant solution in Eq.
(\ref{collapse})
in approximately
half of the parameter-space $\left(\alpha , \beta\right)$. \\

Finally, let us also compute the probability $f$ of string collapse
in the region {\bf I} of parameter space:
\begin{eqnarray}
{f} = \int_{R\leq R_S}^{}d\alpha\,d\beta
\end{eqnarray}
where $R_{S}=4\pi A G\mu\;$ is the Schwarzschild radius of the
string. Using Eqs. (8)-(11), and assuming that $G\mu<<1$
\cite{shellard},  one finds \cite{rene}:
\begin{eqnarray}
{f} = \frac{12\sqrt{6}}{5}\left(4\pi G\mu\right)^{\frac{5}{2}}
\int_{0}^{1} \frac{t^{2}dt}{\sqrt{1-
t^4}}\,+\,\mathcal{O}\left(\left(G\mu\right)^{\frac{7}{2}}\right)
\nonumber \\ =
\frac{3^{\frac{3}{2}}\left(4\pi\right)^4}{5\,
\Gamma^{2}\left(\frac{1}{4}\right)}
\left(G\mu\right)^{\frac{5}{2}}\,+\,\mathcal{O}
\left(\left(G\mu\right)^{\frac{7}
{2}}\right)
\label{approxprob}
\end{eqnarray}
The result (\ref{approxprob}) is a very good approximation
for $G\mu < 10^{-2}$, thus
for any ``realistic" cosmic strings we conclude:
\begin{equation}
{f} \approx 2\cdot10^{3}\cdot\left(G\mu\right)^{\frac{5}{2}}
\end{equation}
Our result (13) partly agrees with that of Ref. \cite{pol} in
the sense that
$f\propto(G\mu)^{5/2}$. However, we find that there is in addition
 a large numerical prefactor in the relation. This factor
is of the order $10^3$. \\

To conclude, simple explicit examples show that the conclusions of
\cite{pol} concerning the minimal string size of the Turok-strings
are generally not correct. In this comment we  re-analyzed the
problem and performed a classification of the Turok-strings, to
clarify the situation. We also computed the probability of string
collapse again, and found that the original result \cite{pol} is off
by approximately 3 orders of magnitude. \\

\begin{figure}[htb]
\vspace{-6cm}
\centerline{\psfig{file=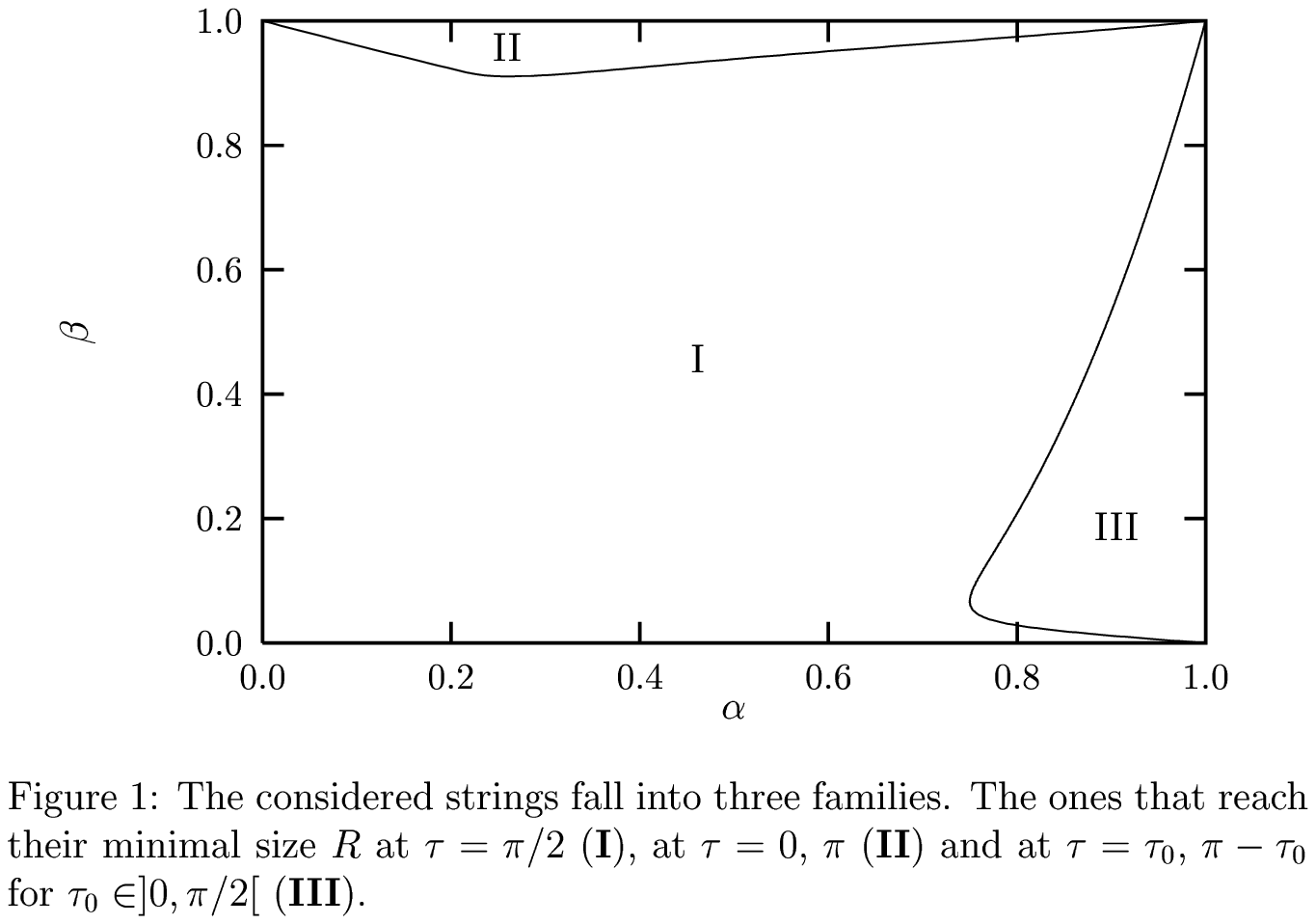,height=29cm}}
\label{fig1}
\end{figure}

\end{document}